\documentclass[useAMS,usenatbib,twocolumn,preprintnumbers,footinbib]{revtex4}
\usepackage{epsfig}
\input psfig.sty

\usepackage{graphicx}
\usepackage{dcolumn}
\usepackage{bm}
\usepackage[english]{babel}

\newcommand{\mytilde}{\raise.17ex\hbox{$\scriptstyle\mathtt{\sim}$}}

\newcommand{\aap}{Astron. and Astrophys.}

\newcommand{\no}{\mbox{$n_{e_o}$}}

\newcommand{\dTo}{\mbox{$\Delta T_0$}}

\newcommand{\sigT}{\mbox{$\sigma_{\mbox{\tiny T}}$}}
\newcommand{\Tcmb}{\mbox{$T_{\mbox{\tiny CMB}}$}}

\newcommand{\Da}{\mbox{$D_{\!\mbox{\tiny A}}$}}

\begin{document}

\title{Constraints on a possible variation of the fine structure constant\\ from galaxy cluster data}

\author{R. F. L. Holanda$^{1,2}$\footnote{E-mail: holanda@uepb.edu.br}}

\author{S. J. Landau$^3$\footnote{E-mail: slandau@df.uba.ar }}

\author{J. S. Alcaniz$^4$\footnote{E-mail: alcaniz@on.br}}

\author{I. E. S\'anchez G.$^3$\footnote{E-mail: isg.cos@gmail.com }}

\author{ V. C. Busti$^5$\footnote{E-mail: vinicius.busti@astro.iag.usp.br}}

\address{$^1$Departamento de F\'{\i}sica, Universidade Estadual da Para\'{\i}ba, 58429-500, Campina Grande - PB, Brasil}

\address{$^2$Departamento de F\'{\i}sica, Universidade Federal de Campina Grande, 58429-900, Campina Grande - PB, Brasil}

\address{$^3$Departamento de F\'{\i}sica, Facultad de Ciencias Exactas y Naturales, Universidad de Buenos Aires\\ and IFIBA, CONICET, Ciudad Universitaria - PabI, Buenos Aires 1428, Argentina}

\address{$^4$Departamento de Astronomia, Observat\'orio Nacional, 20921-400, Rio de Janeiro - RJ, Brasil}

\address{$^5$
Departamento de F\'{\i}sica Matem\'{a}tica, Instituto de F\'{\i}sica, Universidade de S\~{a}o Paulo, 
CP 66318, 05508-090, S\~{a}o Paulo - SP, Brasil}

\date{\today}

\begin{abstract}
We propose a new method to probe a possible time evolution of the fine structure constant $\alpha$ from X-ray and Sunyaev-Zeldovich measurements of the gas mass fraction ($f_{gas}$) in galaxy clusters. Taking into account a direct relation between variations of $\alpha$ and violations of the distance-duality relation, we discuss constraints on $\alpha$ for a class of dilaton runaway models. Although not yet competitive with bounds from high-$z$ quasar absorption systems, our constraints, considering a sample of 29 measurements of $f_{gas}$, in the redshift interval $0.14 < z < 0.89$, provide an independent estimate of $\alpha$ variation at low and intermediate redshifts. Furthermore, current and planned surveys will provide a larger amount of  data and thus allow to improve the limits on $\alpha$ variation obtained in the present analysis. 
\end{abstract}

\pacs{98.80.-k, 98.80.Es, 98.65.Cw}

\maketitle
\section{Introduction}
Since the large number hypothesis proposed by Dirac~\cite{Dirac37}, a possible time variation of fundamental constants has motivated numerous theoretical and experimental investigations.  Conversely, the most commonly accepted cosmological theories are based on the assumption that fundamental constants like the gravitational constant $G$, the fine structure constant $\alpha$, or the proton-to-electron mass ratio $\mu = \frac{m_p}{m_e}$ are indeed truly and genuinely constant. Therefore, the assumption that  these constants do not vary with time or location is just a hypothesis, which needs to be corroborated with observational and experimental data. In fact, several  grand-unification theories predict that these constants are slowly varying functions of low-mass dynamical scalar fields (see \cite{Uzan03} and references therein). In particular, the attempt to  unify all fundamental interactions resulted in the development of multidimensional theories like string-motivated field theories, related brane-world 
theories, and (
related or not) Kaluza-Klein theories which predict not only energy dependence of the fundamental constants but also dependence of their low-energy limits on cosmological time. 

Theoretical frameworks based on first principles were developed by different authors to study the variation of the fine structure constant. For example, string theory models predicts the dilaton field, denoted by $\phi$, as a scalar partner of the spin-2 graviton. The vacuum expectation value of the dilaton determines the string coupling constant $g_s=\exp \phi/2$. In the dilaton scenario studied by \cite{DPV2002a}, the runaway of the field towards strong coupling may yield variations of the fine-structure constant.

In the last decade the issue of the variation of fundamental constants has experienced a renewed interest, and several observational analyses have been performed to study  their possible variations \cite{Uzan2011,GarciaBerro07} and to establish bounds on such variations. The experimental research can be grouped into astronomical and local methods. { {The last ones provide the most stringent bounds in $\alpha$ variation: i) the Oklo natural nuclear reactor that operated about $1.8\times 10^9$ years ago\cite{DD96,Petrov06,Gould06,Onegin12}  yields $-0.7 \times 10^{-8} < \frac{\Delta \alpha}{\alpha} < 10^{-8}$ and ii) laboratory measurements  of atomic clocks with different atomic numbers \cite{Fischer04,Peik04,Rosenband2008} which provide the most strict bound: $\frac{\Delta \alpha}{\alpha} = 1.6 \pm 2.3 \times 10^{-17} $. On the other hand, experiments that test the Weak Equivalence Principle such as torsion balance experiments \cite{Adelberger2009,torsion} and the Lunar Laser Ranging \cite{LLR} can also give 
indirect bounds on the present spatial variation of $\alpha$ \cite{Kraiselburd2012} or on the present time variation of $\alpha$ \cite{Uzan2011} \footnote{In the latter case, the bounds on  $\alpha$ variation obtained from limits on the WEP are model dependent, since the calculation requires to fix  free parameters of the theoretical model for $\alpha$ variation.}. }}

The astronomical methods are based mainly on the analysis of high-redshift quasar absorption systems. In particular, the most successful method employed so far to measure possible variations of $\alpha$ (the so-called many-multiplet method) compares the characteristics of different transitions in the same absorption cloud, and results in a gain of an order of magnitude in sensibility with respect to previous methods  \cite{Webb99}. Most of the reported results are consistent with a null variation of fundamental constants. Nevertheless, Murphy et al.  \cite{Murphy03,King12} have reported results which suggest  a possible spatial variation in $\alpha$ using Keck/HIRES and VLT/UVES observations. However a recent analysis  of the instrumental systematic errors \cite{Whitmore14} of the VLT/UVES data shows that there is no evidence for a space or time variation in $\alpha$ from quasar data. On the other hand, constraints on the variation of $\alpha$ in the early universe can be investigated by using the available 
Cosmic Microwave Background (CMB) data \cite{obrian15,Planckalfa14} and the abundances of the light elements generated during the Big Bang Nucleosynthesis \cite{Mosquera2013}. In Ref.~\cite{Galli2013}, the linear relation between the integrated comptonization parameter, $Y_{SZ}D^2$, and its X-ray counterpart $Y_X$, was used to constrain a possible evolution of the fine structure constant. The application of this method to 61 galaxy clusters from a subsample of the Planck Early Sunyaev-Zel’dovich cluster sample placed tight bounds on $\alpha$ in the redshift interval $0.044 < z < 0.44$ and showed no significant time evolution. 

In this paper, we propose a new method to constrain a possible variation of the fine structure constant from galaxy cluster observations. Differently from the analysis of Ref.~\cite{Galli2013},  we use both X-ray and Sunyaev-Zeldovich measurements of the gas mass fraction ($f_{gas}$) of galaxy clusters and take into account  a direct relation, shown in Ref.~\cite{hees}, between  variations of $\alpha$ and violations of the so-called distance-duality relation $D_L(1+z)^2/D_A=1$, where $D_L$ and and $D_A$ are, respectively, the luminosity and angular diameter distance to a given source at redshift $z$. { {It should be noted that all the above mentioned local experiments but the Oklo mine provide bounds  on the present  time or space variation of $\alpha$ while the current work studies variations for $0.14 < z < 0.89$. In that sense, only the Oklo bound should be compared with the results obtained in this paper.
In Sec. II we discuss the theoretical models used in our analysis. The method here proposed is presented in detail in Sec. III and its application, considering a sample of 29 measurements of  $f_{gas}$ ($0.14 < z < 0.89$) analysed in Ref.~\cite{laroque}, is discussed in Sec. IV. We end the paper with a discussion of the main results in Sec. V.}}

\section{Theoretical Models}

In this section we present the theoretical models we will use to estimate the bounds on the time variation of $\alpha$ with galaxy cluster data. 
Any theory in which the local coupling constants become effectively  spacetime dependent, while respecting the principles of locality and  general covariance,   will involve some  kind of fundamental field (usually a scalar field) controlling such dependence. Thus, this scalar field  will couple in different manners to the various types of matter.  This is the reason why most theories that predict variation of fundamental constants, also forecast effective  violations of the Weak Equivalence Principle (WEP). This issue has been extensively studied in the literature (see, e.g., \cite{Bekenstein82,BMS02,Olive02,DP94}). From these analyses, it follows that there are two models that can not be ruled out by experiments that test violations on the WEP: dilaton runaway models \cite{DPV2002a,DPV2002b,Martins2015} and chameleon models \cite{KW04,Brax04,MS07}. Furthermore, in a recent work, Kraiselburd {\it{et al}}~\cite{Kraiselburd} have shown that only some cases of the chameleon 
 model survives the experimental 
constraints. 

In our analysis, we focus on the dilaton runaway models. The basic idea is to exploit the string-loop modifications of the (four dimensional) effective low-energy action where the Lagrangian can be written as: 

\begin{equation}\label{lagr}
{\cal L}=\frac{R}{16\pi G}-\frac{1}{8\pi G}\left(\nabla\phi\right)^2-\frac{1}{4}B_F(\phi)F^2+... \,.
\end{equation}
where $R$ is the Ricci scalar, $\phi$ is a scalar field, namely the dilaton and $B_F$ is the gauge coupling function. From this action, it is straightforward to obtain the Friedmann equation and the equation of motion for the dilaton field: 

\begin{equation}
H^2=8\pi G \frac{\rho}{3+(1+z)\frac{d\phi}{dz}},
\end{equation}
\begin{eqnarray}
(1+z)^2\frac{d^2\phi}{dz^2}+\left[1-\frac{8\pi G}{2H^2}(\rho-p)\right](1+z)\frac{d\phi}{dz}= \nonumber \\
 -\frac{8\pi G}{2H^2} \sum_i \beta_i(\phi) (\rho_i - 3 p_i),
\end{eqnarray}
where $H$ is the Hubble parameter that is related to the components of the universe and the dilaton field,  $\rho=\sum_i\rho_i$ and $p=\sum_i p_i$ are the total energy density and the pressure respectively, except the corresponding part of the dilaton field. The $\beta_i$ are the couplings of the dilaton with each component of matter $i$; generically the dilaton has different couplings to different components. The relevant parameter for studying the variation of $\alpha$ here is the coupling of the dilaton field to hadronic matter. The crucial assumption is that all gauge fields couple to the same $B_F$ and it follows from Eq. (\ref{lagr}) that  $\alpha \sim B_F^{-1}(\phi)$. Thus, we can write:
\begin{equation}\label{evolfull}
\frac{\Delta\alpha}{\alpha}(z)=\frac{1}{40}\beta_{had,0}\left[1-e^{-(\phi(z)-\phi_0)}\right]\,.
\end{equation}
where $\beta_{had,0}$ is the current value of the coupling between the dilaton and hadronic matter and 
\begin{equation}\label{alphahad}
\beta_{had}(\phi)\sim 40 \frac{\partial\ln B_F^{-1}(\phi)}{\partial\phi}\,\sim (1-b_Fe^{-c\phi})\,.
\end{equation}
where $c$ and $b_F$ are constant free parameters.

In the present work we are interested in the evolution of the dilaton at low redsfhits $ 0.14 < z < 0.89$ and thus it is a reasonable approximation to linearize the field evolution. In such way, we obtain the following expression:
\begin{equation}\label{evolslow}
\frac{\Delta\alpha}{\alpha}(z)\approx\, -\frac{1}{40}\beta_{had,0} {\phi_0'}\ln{(1+z)}\,;
\end{equation}
where $\phi_0'= \frac{\partial \phi}{\partial \ln a}$ at present time. This last equation is the one we will use to compare the model predictions with galaxy cluster data through the method discussed in the next section.

\section{Method}

Observations of the gas mass fraction in relaxed and massive galaxy clusters have been widely used as a cosmological test (see, e.g., [7-13] and references therein). The gas mass fraction is defined as \cite{sasaki}
\begin{equation}
 f_{gas}=\frac{M_{gas}}{M_{tot}},
 \label{eq3.14}
\end{equation}
where $M_{tot}$ is the total mass  and  $M_{gas}$ is the gas mass obtained by  integrating the gas density model. In spherical coordinates, it is written as
\begin{eqnarray}
 M_{gas}(<V)=4\pi\int_{0}^{R}{\rho_{gas}r^2dr}.
 \label{eq3.4}
\end{eqnarray}
The intracluster gas comes from the primordial gas and we can consider that it consists only of hydrogen (H) and helium ($H_e$). Thus,
\begin{eqnarray}
 n_H=\left(\frac{2X}{1+X}\right)n_e(r)
 \label{eq3.6}
\end{eqnarray}
and
\begin{eqnarray}
 n_{He}=\left[\frac{1-X}{2(1+X)}\right]n_e(r),
 \label{eq3.7}
\end{eqnarray}
where $X$ is the hydrogen abundance and

\begin{eqnarray}
 \rho_{gas}&=&\rho_H+\rho_{He} \nonumber \\
 \rho_{gas}&=&\frac{2n_{e0}m_H}{(1+X)}\left(1+\frac{r^2}{r_{c}^{2}}\right)^{-\frac{3\beta}{2}},
 \label{eq3.8}
\end{eqnarray}
with $m_H$ being the hydrogen mass and $r_c$  the core radius. In the above equation we also have used the isothermal spherical $\beta$ model to describe the electronic density \cite{cavaliere}
\begin{equation}
n_e({\mathbf{r}}) = \no \left (1 + \frac{r^2}{r_c^2} \right )^{-3\beta/2}.
\label{eq:single_beta}
\end{equation}
From the above equations, we obtain

\begin{eqnarray}
 M_{gas}(<V)=\frac{8\pi n_{e0}m_H}{(1+X)}\int_{0}^{R}{\left(1+\frac{r^2}{r_{c}^{2}}\right)^{-\frac{3\beta}{2}} r^2dr}\;,
 \label{eq3.9}
\end{eqnarray}
or still
\begin{eqnarray}
 M_{gas}(<V)=\frac{8\pi n_{e0}m_Hr_{c}^{3}}{(1+X)}I_M(y,\beta),
 \label{Mgas}
\end{eqnarray}
where
\begin{eqnarray}
 I_M(R/r_c,\beta)\equiv \int_{0}^{R/r_c}{(1+x^2)^{-\frac{3\beta}{2}}x^2dx}\;,
 \label{eq3.10}
\end{eqnarray}
and $x=r/r_c$.

On the other hand, under the assumption of hydrostatic equilibrium assumption, isothermality and Eq. (\ref{eq:single_beta}),   $M_{tot}$ is given by \cite{grego}
\begin{eqnarray}
 M_{tot}(<R)=\frac{3\beta k_BT_G}{\mu Gm_H}\left[\frac{R^3}{(r_{c}^{2}+R^2)}\right],
 \label{Mtot}
\end{eqnarray}
where $T_{G}$ is the temperature of the intracluster medium obtained from X-ray spectrum, $\mu$ and $m_p$ are, respectively, the total mean molecular weight and the proton mass, $k_{\rm B}$ the Boltzmann constant and $G$ is the gravitational constant. 

Finally, by using equations (\ref{Mgas}) and  (\ref{Mtot}) we have

\begin{eqnarray}
f_{gas}=\frac{8\pi m_H^2\mu G n_{e0}}{3(1+X)\beta k_BT_G}\left[\frac{(r_{c}^{5}+r_c^3R^2)}{R^3}\right]I_M({R \over r_c},\beta).
\label{fgas} 
\end{eqnarray}
The parameter $n_ {e0}$ in the above equation can be determined from two different kinds of observations: X-rays surface brightness and the Sunyaev-Zeldovich effect. In what follows, we discuss these observations and make explicit the $f_{gas}$ dependence with respect to the $\alpha$ parameter.

\subsection{X-ray Observations}

At high temperatures, the intergalactic gas emits mainly through thermal bremsstrahlung (see, e.g., \cite{sarazin}). The bolometric luminosity is given by
\begin{eqnarray}
 L_x=4\pi\int_{0}^{R}{\frac{dL_x}{dV}r^2dr},
 \label{eq3.16}
\end{eqnarray}
with
\begin{eqnarray}
 \frac{dL_x}{dV}=\left(\frac{2\pi k_BT_G}{3m_e}\right)^{\frac{1}{2}}\frac{2^4e^6}{3\hbar m_ec^3}n_e\left(\sum_i{Z_i^2n_ig_{Bi}}\right),
 \label{eq3.17}
\end{eqnarray}
where $m_e$ is the electron mass, $e$ is the electronic charge, $\hbar$ is the Planck constant divided by $2\pi$, $c$ is the speed of light, $n_e$ is the electronic density of gas and $Z_i$ and $n_i$ are, respectively, the atomic numbers and the distribution of elements. $g_B$ is the Gaunt factor which takes into account the corrections due quantum and relativistic  effects of Bremsstrahlung emission. Again, by considering the intracluster medium constituted  by  hydrogen and helium and the spherical $\beta$ model we have

\begin{eqnarray}
 L_x&=&\left(\frac{2\pi k_BT_G}{3m_e}\right)^{\frac{1}{2}}\frac{2^4e^6}{3\hbar m_ec^3}g_B(T_G)\frac{2}{(1+X)}4\pi n_{e0}\nonumber \\
 &\times& \int_{0}^{R}{\left(1+\frac{r^2}{r_{c}^{2}}\right)^{-3\beta}r^2dr}.
 \label{eq3.19}
\end{eqnarray}
Now, defining
\begin{eqnarray}
I_L(R/r_c,\beta)\equiv \int_{0}^{R/r_c}{(1+x^2)^{-3\beta}x^2dx}\;,
\label{eq3.20}
\end{eqnarray}
where $x=r/r_c$, we obtain the equation for the bolometric luminosity
\begin{eqnarray}
 L_x&=&\left(\frac{2\pi k_BT_G}{3m_e}\right)^{\frac{1}{2}}\frac{2^4e^6}{3\hbar m_ec^3}g_B(T_G)\frac{2}{(1+X)}4\pi n_{e0}^2r_{c}^{3}\nonumber \\
 &\times& I_L(R/r_c,\beta),
 \label{eq3.21}
\end{eqnarray}
which can be rewritten as 
\begin{eqnarray}
 L_x&=&\alpha^3\left(\frac{2\pi k_BT_G}{3m_e}\right)^{\frac{1}{2}}\frac{2^4\hbar^2}{3 m_e}g_B(T_G)\frac{2}{(1+X)}4\pi n_{e0}^2 D^2_A \theta^2_c r_{c}\nonumber \\
 &\times& I_L(R/r_c,\beta)\;,
 \label{eq3.22}
\end{eqnarray}
{  where  $D_A$ is the angular diameter distance. The quantity $L_x$, the total X-ray energy per second leaving the galaxy cluster, is not an observable. The observable is the X-ray flux, given by
\begin{equation}
 F^x = L_x/4\pi D^2_L,
\label{fluxo}
\end{equation}
where $D_L$ is the luminosity distance. Thus, it is possible to see from Eq. (\ref{fluxo}) that $n_{e0}$  is  $\propto \alpha^{-3/2} D_L/D_A$. Therefore, if $\alpha=\alpha_0\varphi(z)$ and the cosmic distance duality relation is $D_L (1+z)^{-2}/D_A=\eta$ (see \cite{holanda}), the gas mass fraction measurements  extracted from X-ray data are affected by a possible departure of $\alpha_0$ and $\eta =1$, such as}
\begin{equation}
f_{X-ray}^{th} \propto  [\varphi(z)]^{-3/2} \eta(z) \;.
\end{equation}

\subsection{Sunyaev-Zel'dovich Observations}\label{sze}

The measured temperature decrement $\Delta T_{\rm SZE}$ of the CMB due to the Sunyaev-Zel'dovich effect \cite{sunyaev} is given by \cite{laroque}
\begin{equation}
\label{eq:sze1} \frac{\Delta T_{\rm 0}}{T_{\rm CMB}} = f(\nu,
T_{\rm e}) \frac{ \sigma_{\rm T} k_{\rm B} }{m_{\rm e} c^2} \int n_e T_{\rm e} dl, \
\end{equation}
where $T_{\rm CMB} =2.728$ K
is the present-day temperature of the CMB, $m_{\rm e}$ the electron mass and $f(\nu, T_{\rm e})$ accounts for frequency shift and relativistic corrections \cite{Itoh} and 
$\sigma_T={8\pi \hbar^2 \alpha^2}/{3 m_e^2 c^2}$ 
 is the Thompson cross section.

Using Eq. (\ref{eq:single_beta}), the central electron density can now be expressed as
\begin{equation}
n_{e0}^{SZE} =  \frac{\dTo \,m_e c^2 \:\Gamma(\frac{3}{2}\beta)}{f_{(\nu, T_e)}
  \Tcmb \sigT\, k_{\rm B} T_e \Da \pi^{1/2} \:\Gamma(\frac{3}{2}\beta -
  \frac{1}{2})\, \theta_c}.
\label{eq:sz_ne0}
\end{equation}
By using the expression for the Thompson scattering cross section, it is straightforward to show that the current gas mass fraction measurements via SZE depend on $\alpha$ as  
\begin{equation}
f_{SZE} \propto {\alpha^{-2} }\;,
\end{equation}
or still 
\begin{equation}
f_{SZE}^{th} \propto  [\varphi(z)]^{-2} \;.
\end{equation}
In our analysis, $\varphi(z)=1-\frac{1}{40}\beta_{had,0} {\phi_0'}\ln{(1+z)}$ [see Eq. (\ref{evolslow})].

\subsection{$f_{SZE}/f_{X-ray}$ relation}

Current $f_{X-ray}$ measurements have been obtained by assuming $\varphi(z)=1$ and $\eta=1$. If, however, $\alpha$ varies over the cosmic time, the real gas mass fraction from X-ray ($f_{X-ray}^{th}$) and SZE ($f_{SZE}^{th}$) observations  should be related with the current  observations by 
\begin{equation}
f_{X-ray}^{th}=\varphi(z)^{-3/2} \eta(z) f_{X-ray}\;,
\end{equation}
\begin{equation}
f_{SZE}^{th}=\varphi(z)^{-2} f_{SZE}\;. 
\end{equation}
Thus, if all the physics behind the X-ray and SZE observations are properly taken into account, one would expect $f_{gas}$ measurements from both techniques to agree with each other since they are measuring the very same physical quantity. Therefore, considering only the variation of $\alpha$, the expression relating current X-ray  and  SZE  observations is given by:
\begin{equation} \label{relation}
f_{SZE}=\varphi(z)^{1/2} \eta(z) f_{X-ray}\;.
\end{equation}

{  Before proceeding further with the observational analysis, an important aspect should be considered. As shown in Ref.~\cite{hees}, variations of the fine structure constant and violations of the so-called distance-duality relation $D_L(1+z)^2/D_A=1$, where $D_L$ is the luminosity distance and $D_A$ is the angular diameter distance, are intimately and unequivocally linked. In particular, rewriting the latter expression as  $D_L(1+z)^2/D_A=\eta$, where $\eta$ quantifies possible departures from the cosmic distance duality relation, these authors showed that constraints on the function $\eta(z)$ can be translated into constraints on the temporal variation of $\alpha$ as
\begin{equation}\label{eq:dalpha_eta}
\frac{\Delta \alpha(z)}{\alpha}=\frac{\alpha(z)-\alpha_0}{\alpha_0}=\frac{h(\phi_0)}{h(\phi)}-1=\eta^2(z)-1.
\end{equation}
 In our case, $\Delta \alpha/\alpha = (\alpha_0 \varphi - \alpha_0)/\alpha_0=\varphi -1$. Therefore, combining the above results from Refs. \cite{hees}, our final expression can be written as}
\begin{equation} \label{relation}
f_{SZE}=\eta(z)\varphi(z)^{1/2} f_{X-ray}\;,
\end{equation}
or, equivalently,
\begin{equation} 
f_{SZE}=\varphi(z) f_{X-ray}\;,
\label{fsz2}
\end{equation}
which provides a direct test for $\varphi(z)$ taking into account effects from both a possible variation of $\alpha$ as well as a possible violation of the distance duality relation. It is also worth mentioning that, since the above expression holds for a given object, systematic errors on the $\eta$ estimates due to redshift differences of distinct objects  (e.g., in tests involving SNe Ia and galaxy clusters or baryon acoustic oscillation data) are fully removed.

\begin{figure}[t]
\mbox{\hspace{-0.8cm}
\includegraphics[width = 3.2in, height = 3.0in]{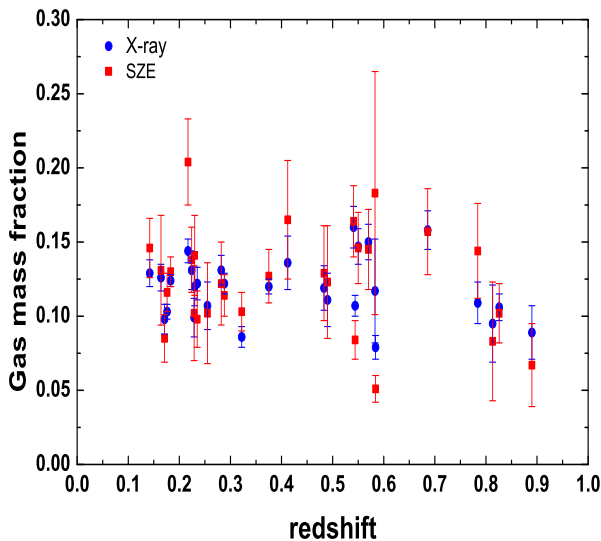}}
\caption{Gas mass fraction measurements as a function of redshift considered in our analysis (subsample of 29 galaxy clusters) \cite{laroque}. Blue circles and red squares stand for measurements obtained from X-ray and SZE observations, respectively.}
\label{Fig}
\end{figure}

\section{analysis and results}

Recently, La Roque {\it{et al.}} \cite{laroque} provided a sample of measurements of the gas mass fraction of galaxy clusters from both X-ray and SZ observations. 
This sample, derived from Chandra X-ray observations and OVRO/BIMA  interferometric Sunyaev-Zel'dovich Effect (SZE) measurements, comprises 38 $f_{gas}$ data points  in the redshift interval $0.14 \leq z \leq 0.89$. In their analysis, the authors used the nonisothermal double $\beta$-model for gas distribution and the 3D temperature profile was modelled assuming that the ICM is in hydrostatic equilibrium with a NFW dark matter density distribution \cite{navarro}. The gas density model was obtained from a joint analysis of the X-ray and SZE data, which makes the SZE gas mass fraction not completely independent. However, Ref. ~\cite{laroque} used $r_{2500}$ (the radius at which the mean enclosed mass density is equal to $2500$ cosmological critical density) in their analysis and  current simulations \cite{halmann}  have shown  that the shape parameter values computed 
separately by SZE and X-ray observations agrees at $1\sigma$ level within this radius. 

When described by the hydrostatic equilibrium model, some objects of the La Roque {\it{et al.}} sample presented questionable  reduced $\chi^{2}$ ($2.43 \leq \chi_{d.o.f.}^2 \leq 41.62$).  They are: Abell 665, ZW 3146, RX J1347.5-1145, MS 1358.4 + 6245, Abell 1835, MACS J1423+2404, Abell 1914, Abell 2163, Abell 2204. By excluding these objects from our analysis we end up with a subsample of 29 galaxy clusters, shown in Figure 1.

We evaluate our statistical analysis by defining the likelihood distribution function ${\cal{L}} \propto e^{-\chi^{2}/2}$, where
\begin{equation}
\label{chi2} \chi^{2} = \sum_{i = 1}^{N}\frac{{\left[1-\frac{1}{40}\beta_{had,0} {\phi_0'}\ln{(1+z)} - \phi_{i, obs}(z) \right] }^{2}}{\sigma^{2}_{i, obs}},
\end{equation}
{$\phi_{i, obs}(z) = f_{SZE}/f_{X-ray}$ and $\sigma^{2}_{i, obs}$ is the uncertainty }associated to this quantity, i. e.,

\begin{equation}
\sigma^{2}_{i, obs} =\left[\frac{1}
{f_{X-ray}}\right]^2 \sigma^2_{f_{SZE}}+ \nonumber \\
 \left[\frac{f_{SZE}}
{(f_{X-ray})^2}\right]^2\sigma^2_{f_{X-ray}}.
\end{equation}
{  The error bars of gas mass fraction measurements  take into account the statistical errors of the X-ray and SZE observations  estimated in Ref.~\cite{laroque}. The common statistical contributions to $f_{gas}$ are: SZE point sources $\pm 4\%$,  kinetic SZ $\pm 8\%$, $\pm 20\%$  and $\pm 10\%$ for cluster asphericity to $f_{gas}$ from X-ray and SZE observations, respectively. The asymmetric error bars were treated as discussed in \cite{Ddah}, i. e., $f_{gas} =  \widetilde{f}_{gas} + \Delta_+ - \Delta_-$, with $\sigma_{f_{gas}} = (\Delta_+ +\Delta_-)/2$, where  $\widetilde{f}_{gas}$ stands for the La Roque et al. (2006) measurements and $\Delta_+$ and $\Delta_-$ are, respectively, the associated upper and lower errors bars. On the other hand, the systematic errors for the galaxy clusters are:  X-ray absolute flux calibration $\pm 6\%$, X-ray temperature calibration $\pm 7.5\%$, SZE calibration $\pm 8\%$  and a one-sided systematic uncertainty of $-10\%$ to the total masses, which accounts for 
the assumed hydrostatic equilibrium. We also performed our analysis by combining the statistical and systematic errors in quadrature for the gas mass fractions of galaxy clusters.}

Constraints on the quantity $\gamma = \frac{1}{40}\beta_{had,0} {\phi_0'}$ are shown in Figure 2. Solid and dashed curves correspond to analyses with and without systematic errors, respectively. We obtain $\gamma = 0.037\pm 0.18$ and $\gamma = 0.065\pm 0.095$ at 68.3\% (C.L.), which are fully compatible with $\phi(z)=1$ or, equivalently, with no variation of fine structure constant $\alpha$.
Now, it is interesting to compare our bound on $\gamma $ with limits obtained using independent data. From the expression for the deceleration parameter $q_0$ and considering the Planck estimate for the total matter density $\Omega_m$, an upper bound on the value of $\phi_0'$ can be estimated, namely, $\phi_0'\leq 0.3$ at 1$\sigma$ c.l. \cite{Martins2015}. Torsion balance tests and lunar laser ranging provide $\beta_{had,0} \leq 10^{-4}$ \cite{torsion,LLR} whereas a comparison of atomic clocks with differente atomic number implies $\gamma \leq 3 \times 10^{-5}$~\cite{Rosenband2008}.  It follows from this discussion, that the bounds obtained in this work from current $f_{gas}$ data, although less stringent, are fully consistent with limits obtained with independent data.

\begin{figure}[t]
\mbox{\hspace{-0.8cm}
\includegraphics[width = 3.2in, height = 3.0in]{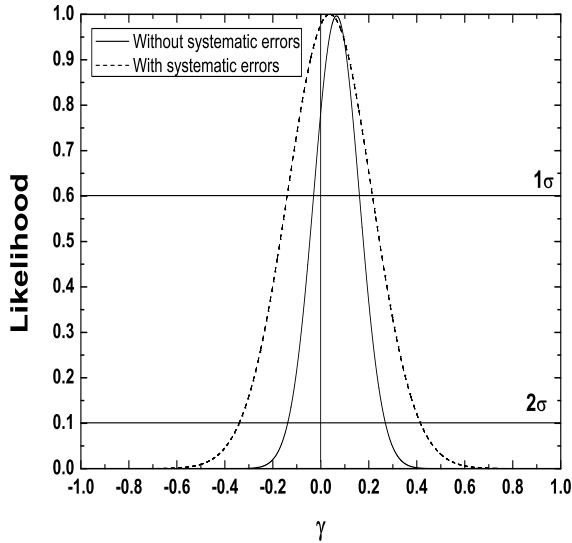}}
\caption{Constraints on a possible variation of the fine structure constant. The solid and dashed lines correspond to analyses with and without the systematic errors discussed in the text, respectively.} 
\label{Fig2}
\end{figure}

\section{Conclusions}

The search for a possible variation of fundamental constants constitutes an important task not only for cosmology but 
also for fundamental physics (see, e.g., \cite{lang}). In this work we have proposed a new method to investigate the 
time variation of the fine structure constant $\alpha$ in cosmological scales. We have shown that observations of 
the gas mass fraction via the Sunyaev-Zeldovich effect and X-ray surface brightness of the same galaxy cluster are 
related by $f_{SZE}=\varphi(z)f_{X-ray}$, where $\varphi(z)=\frac{\alpha}{\alpha_0}$, which furnishes constraints on 
the evolution of $\alpha$. 

We have also applied such a method to a sample of 29  measurements of the gas mass fraction of galaxy clusters very 
well described by the hydrostatic equilibrium assumption, as discussed in Ref.~\cite{laroque}. Taking into account 
the results of Ref.~\cite{hees}, connecting variations of $\alpha$ with violation in the duality distance relation, 
we have derived, for a class of dilaton runaway models, new constraints on $\alpha$ which are fully compatible 
with $\phi(z)=1$.  It is worth mentioning that a similar analysis using different galaxy cluster observables has been performed in Ref.~\cite{Galli2013}. Differently from our results, however,  the variation of $\alpha$ induced by the distance-duality relation was not considered in this latter work, which suggests that the bounds on $\alpha$ there derived should be revised. Given the small number of $f_{gas}$ data points currently available, the constraints reported here are not yet competitive with the tight bounds derived from other analises (see, e.g., \cite{hees} and references therein). We believe, however,  that when applied to upcoming galaxy cluster data from current and planned surveys (e.g., SPT~\footnote{http:// pole.uchicago.edu/} and eROSITA~\footnote{http://www.mpe.mpg.de/eROSITA}) the method discussed in this paper may be useful to probe a possible variation of the fine structure constant as well as to explore its theoretical consequences.

\section*{Acknowledgments}

RFLH acknowledges financial support from INCT-A and CNPq (No. 478524/2013-7). SL is  supported  by 
PIP 11220120100504  CONICET. JSA is supported by CNPq, INEspa\c{c}o and Rio de Janeiro Research Foundation (FAPERJ).  VCB is supported by 
S\~{a}o Paulo Research Foundation (FAPESP)/CAPES agreement, under grant number 2014/21098-1.

\bibliography{bibliography}
\bibliographystyle{apsrev}

\end{document}